\documentclass[11pt]{article}
\usepackage{amsfonts,amsmath,graphicx,setspace}
\usepackage{color,hyperref,verbatim}
\usepackage[super, sort&compress]{natbib}

\definecolor{Blue}{rgb}{0,0,0.5}
\hypersetup{%
    colorlinks = {true}, linktocpage = {true},
    plainpages = {false}, linkcolor = {Blue},
    citecolor = {Blue}, urlcolor = {Blue},
    pdfstartview = {XYZ null null 1.25},
    pdfpagemode = {UseOutlines},
    pdfview = {XYZ null null null}
}
\setcitestyle{citesep={,}}

\newcommand{\email}[1]{\href{mailto:#1}{\normalfont\texttt{#1}}}
\newcommand{\pkg}[1]{{\fontseries{b}\selectfont #1}}
\newcommand{\bfalpha}{\mbox{{\boldmath $\alpha$}}}
\newcommand{\bfdelta}{\mbox{{\boldmath $\delta$}}}
\newcommand{\bfbeta}{\mbox{{\boldmath $\beta$}}}
\newcommand{\bfgamma}{\mbox{{\boldmath $\gamma$}}}
\newcommand{\bflambda}{\mbox{{\boldmath $\lambda$}}}
\newcommand{\bftheta}{\mbox{{\boldmath $\theta$}}}
\newcommand{\bfsigma}{\mbox{{\boldmath $\sigma$}}}
\newcommand{\bfxi}{\mbox{{\boldmath $\xi$}}}
\newcommand{\bfpsi}{\mbox{{\boldmath $\psi$}}}
\newcommand{\bfOmega}{\mbox{{\boldmath $\Omega$}}}
\newcommand{\eps}{\varepsilon}
\newcommand{\bw}{\mbox{{\boldmath $w$}}}
\newcommand{\bv}{\mbox{{\boldmath $v$}}}
\newcommand{\by}{\mbox{{\boldmath $y$}}}
\newcommand{\bY}{\mbox{{\boldmath $Y$}}}
\newcommand{\bT}{\mbox{{\boldmath $T$}}}
\newcommand{\bx}{\mbox{{\boldmath $x$}}}
\newcommand{\bX}{\mbox{{\boldmath $X$}}}
\newcommand{\bI}{\mbox{{\boldmath $I$}}}
\newcommand{\bbI}{\mbox{\textbf I}}
\newcommand{\dd}{\mbox{d}}
\newcommand{\bz}{\mbox{{\boldmath $z$}}}
\newcommand{\bZ}{\mbox{{\boldmath $Z$}}}
\newcommand{\bK}{\mbox{{\boldmath $K$}}}
\newcommand{\bN}{\mbox{{\boldmath $N$}}}
\newcommand{\bb}{\mbox{{\boldmath $b$}}}
\newcommand{\bu}{\mbox{{\boldmath $u$}}}
\newcommand{\bB}{\mbox{{\boldmath $B$}}}
\newcommand{\bH}{\mbox{{\boldmath $H$}}}
\newcommand{\br}{\mbox{{\boldmath $r$}}}
\newcommand{\bD}{\mbox{{\boldmath $D$}}}
\newcommand{\var}[1]{\mbox{\texttt{#1}}}

\begin{document}

{\vspace*{1cm}}

\begin{center}
\Large \bf Using Joint Models for Longitudinal and Time-to-Event Data to Investigate the Causal Effect of Salvage Therapy after Prostatectomy
\end{center}

\vspace{0.5cm}

\begin{center}
{\large Dimitris Rizopoulos$^{1,2,*}$, Jeremy M.G. Taylor$^{3}$,  Grigorios Papageorgiou$^{1,2}$ and Todd M. Morgan$^{4}$}\footnote{$^*$Correspondence at: Department of Biostatistics, Erasmus University Medical Center, PO Box 2040, 3000 CA Rotterdam, the Netherlands. E-mail address: \email{d.rizopoulos@erasmusmc.nl}.}\\
$^{1}$Department of Biostatistics, Erasmus University Medical Center, the Netherlands\\
$^{2}$Department of Epidemiology, Erasmus University Medical Center, the Netherlands\\
$^{3}$Department of Biostatistics, University of Michigan, Ann Arbor, USA\\
$^{4}$Department of Urology, University of Michigan, Ann Arbor, USA\\
\end{center}

\begin{spacing}{1}
\noindent {\bf Abstract}\\
Prostate cancer patients who undergo prostatectomy are closely monitored for recurrence and metastasis using routine prostate-specific antigen (PSA) measurements. When PSA levels rise, salvage therapies are recommended in order to decrease the risk of metastasis. However, due to the side effects of these therapies and to avoid over-treatment, it is important to understand which patients and when to initiate these salvage therapies. In this work, we use the University of Michigan Prostatectomy registry Data to tackle this question. Due to the observational nature of this data, we face the challenge that PSA is simultaneously a time-varying confounder and an intermediate variable for salvage therapy. We define different causal salvage therapy effects defined conditionally on different specifications of the longitudinal PSA history. We then illustrate how these effects can be estimated using the framework of joint models for longitudinal and time-to-event data. All proposed methodology is implemented in the freely-available \textsf{R} package \textbf{JMbayes2}.
\end{spacing}

\vspace{0.8cm}

\section{Introduction} \label{s:intro}
Many prostate cancer (PCa) patients undergo surgical removal of the prostate gland (radical prostatectomy) as their initial treatment, after being diagnosed with prostate cancer. Even though this surgical procedure is generally successful, the risk of recurrence and metastasis remains. For this reason, urologists closely monitor the prostate-specific antigen (PSA) levels of these patients. PSA can be easily measured from a blood sample, and a persistent rise in the PSA values suggests the cancer may be regrowing, although it is generally not yet detectable on imaging. After the initial surgery, PSA levels drop to near zero; however, PSA may rise again for some patients, leading the treating physicians to recommend salvage therapy to reduce their risk of metastasis. Salvage therapy typically consists of radiotherapy with or without androgen deprivation therapy (ADT), although in some cases ADT alone may be utilized. After salvage therapy, PSA levels nearly always drop, sometimes substantially, but typically rise again if metastasis is going to occur. Because of the serious side effects of the aforementioned salvage therapies and to avoid over-treatment, it is critical to understand which patients are most likely to benefit from these treatments and when they should be initiated. To our knowledge, there are no randomized trials that are designed explicitly to address this question. However, there are clinical trials that compare different types of salvage therapy.\citep{spratt.et.al:18} In this paper, we will not be considering that different salvage therapies have different effects.

We will use the University of Michigan Prostatectomy registry (UMP) Data to tackle this question. This database includes 3634 PCa patients who underwent a radical prostatectomy during the period 1996--2013. Of those patients, 271 (15.6\%) received salvage therapy, 102 (2.8\%) developed metastasis, and 209 (5.8\%) died without metastases. Of these 209 patients, 190 died before salvage therapy, and 19 died after salvage therapy without developing metastasis. For the patients who received salvage, the median PSA just prior to the initiation of the therapy was 0.7 ng/mL (min: 0.0009, Q1: 0.4, Q3: 2.06, max: 266 ng/mL), and 55 had a metastasis. For the patients who did not receive salvage therapy or experience metastases, the median PSA at their last measurement was 0.001 ng/mL (min: 0.0009, Q1: 0.001, Q3: 0.09, max: 11.7 ng/mL). Since death due to prostate cancer is extremely rare without prior metastases, these deaths are considered as due to other causes. Web Figures~1 and 2 show the cumulative incidence functions for metastasis and death and initiation of salvage therapy and death. This dataset has been previously described in \citet{beesley.et.al:19}. The detection of metastasis usually requires imaging. Although imaging is non-invasive, it is not routinely performed without an indication, and the decision to undergo imaging would usually be based on whether the patients exhibited symptoms or if the PSA levels were consistently rising and had attained higher values. There is variation in the schedule of when PSA measurements are taken after the initial surgery, but an average schedule might be every three months for the first year, every six months for the next two years, and annually thereafter. PSA measurements are also collected after salvage therapy is initiated. We aim to utilize the longitudinal PSA measurements and baseline information (Gleason score, T-stage of the tumors, age, race, comorbidities) to quantify the effect of salvage therapy on reducing or delaying metastases and aid the decision-making process of urologists. Given the observational nature of the UMP data, we face several challenges in achieving our goal. First, the longitudinal PSA process affects both metastasis and the assignment of salvage therapy, and therefore it is a time-varying confounder. At the same time, the salvage therapy affects future PSA measurements, and hence the longitudinal PSA process will also be an intermediate variable. In addition, patients who received salvage therapy are more closely monitored, and they will tend to have more PSA measurements and more checks done to see if metastasis has occurred than patients who did not. Finally, death is a competing risk for metastasis and requires accounting for it appropriately.

Nowadays, there is a considerable body of literature on methods developed to estimate causal treatment effects in settings with time-varying treatments when there exists confounding by time-dependent covariates affected by earlier treatments. While matching methods\citep{schaubel.et.al:06, schaubel.et.al:09} and joint modeling approaches\citep{kennedy.et.al:10, taylor.et.al:14} have been developed, most of this literature has focused on marginal structural models, dynamic treatment regimes and related methods. Excellent overviews of these methods are given in \citet{hernan.robins:20} and \citet{tsiatis.et.al:20}. The advantage of these methods is that they are primarily non-parametric, making very few assumptions for the time-varying confounders. However, most of them require specifying a single model for all patients that relates the decision to give treatment on past confounder values. This is especially challenging in our setting because doctors decide when to initiate therapy on different grounds. For example, one physician may use just the last observed value of PSA. In contrast, another may treat patients who showed a sudden increase/decrease in the biomarker in the previous two/three measurements. In situations like this, specifying a single model for the decision to start therapy conditional on past PSA values is challenging. An alternative approach to estimating a causal effect is through direct modeling of the outcome process.\citep{kennedy.et.al:10, xu.et.al:16} We will adopt this approach and avoid having to model the treatment initiation process, and employ the framework of joint models for longitudinal and time-to-event data.\citep{rizopoulos:12} In particular, we postulate a linear mixed-effects model for the longitudinal PSA levels that explicitly accounts for the change in the subject-specific trajectories after initiating salvage therapy. For the instantaneous risk of metastasis, we specify a relative risk model that includes the time-varying salvage therapy and PSA effects, and for the hazard of death, we only include the time-varying salvage therapy effect. Based on the postulated joint model, and for each patient at risk at time $t$, we predict the cumulative risk of metastasis under the two treatment scenarios, initiating salvage therapy at $t$ or continuing without it. Then, we obtain the causal salvage therapy effect by suitably averaging these predicted cumulative risks over appropriate groups of subjects. The benefit of our approach is that it does not require defining a model for initiating treatment.

The rest of the paper is organized as follows. Section~\ref{s:STeffects} presents the definition of causal salvage therapy effects we wish to estimate and the assumptions we make to identify these effects from observational data. Section~\ref{s:Modeling} presents the modeling framework we use to analyze the data, and Section~\ref{s:STeffects_Est} shows the procedure to estimate the salvage therapy effects and derive their variance. In Section~\ref{s:UMPD_Analysis} we apply our proposed methods in the UMP dataset. Finally, Section~\ref{s:Simulation} includes  the results of a simulation study, and Section~\ref{s:Discuss} concludes the paper.


\section{Salvage Therapy Effects} \label{s:STeffects}
\subsection{Definitions} \label{s:STeffects_defs}
We let $T^m$ denote the time to metastasis and $T^d$ the time to death. PSA measurements during follow-up are denoted as $Y(t_j)$ taken at times points $t_j$ ($j = 1, \ldots, J$), and $\mathcal Y(t) = \{Y(t_j) = y(t_j); 0 \leq t_j \leq t, j = 1, \ldots, J\}$ denotes the available PSA measurements history up to time $t$. We denote the time salvage therapy was initiated by $S$, and $N(t) = I(t \geq S)$, $\mathcal N(t) = \{N(t_j); 0 \leq t_j < t, j = 1, \ldots, J\}$ denotes the history of the salvage therapy process. Without the loss of generality, we presume that the urologists' consideration to initiate salvage therapy took place at the same $J$ time points the patients provided PSA measurements. Baseline covariate information is denoted as $\mathcal X = \{X_q = x_q, q = 1, \ldots, Q\}$.

To quantify the causal effect of salvage therapy, we will use the framework of counterfactual outcomes. Namely, we would like to decide if salvage therapy should be initiated at the follow-up time $t$ by comparing the counterfactual cumulative risks of metastasis under the two regimes in the medically-relevant time interval $(t, t + \Delta t]$ conditional on survival, no prior salvage therapy and no metastasis up to $t$. In particular, we let $[T^{m(1)}, \mathcal F^{(1)}(v, t) \mid T^m > t, T^d > t, \mathcal Y(t), \mathcal N(t) = 0, \mathcal X]$ denote the joint conditional distribution of $T^{m(1)} $ and $\mathcal F^{(1)}(v, t)$ for all $v > t$, where $\mathcal F^{(1)}(v, t) = \{Y(t_j); t < t_j \leq v, j = 1, \ldots, J''\}$ denotes the $J''$ future PSA measurements after $t$ and up to the horizon time $v$, if salvage therapy was initiated at $t$ given the available PSA information up to $t$ and the fact that neither metastasis nor death has occurred up to this time. Analogously, we let $[T^{m(0)}, \mathcal F^{(0)}(v, t) \mid T^m > t, T^d > t, \mathcal Y(t), \mathcal N(t) = 0, \mathcal X]$ denote the joint conditional distribution of $T^{m(0)} $ and $\mathcal F^{(0)}(v, t)$ for all $v > t$, if salvage therapy will \emph{not} be initiated in the interval $(t, v]$.

The marginal effect over all possible longitudinal histories up to time $t$ can be defined as:
\begin{eqnarray}
\nonumber \lefteqn{\mbox{ST}^M(t + \Delta t, t) = }\\
\nonumber && \int \biggl [ \Pr \{T^{m(1)} \leq t + \Delta t \mid T^m > t, T^d > t, \mathcal Y(t), \mathcal X\}\\
\nonumber && - \Pr \{T^{m(0)} \leq t + \Delta t \mid T^m > t, T^d > t, \mathcal Y(t), \mathcal X\}\biggr ]\\
\nonumber && \prod \limits_{j = 1}^{J'} \dd G\{Y(t_j), \mathcal X \mid \mathcal Y(t_{j-1})\}\\
\nonumber & = & \Pr \{T^{m(1)} \leq t + \Delta t \mid T^m > t, T^d > t\}\\
& & - \Pr \{T^{m(0)} \leq t + \Delta t \mid T^m > t, T^d > t\}, \label{Eq:mST_def}
\end{eqnarray}
where $G(\cdot)$ denotes the cumulative joint distribution function for $Y$ and $\mathcal X$, and $J'$ denotes the number of longitudinal measurements up to $t$. In this and the following expressions, and for simplicity of exposition, we have omitted that we condition on salvage not being initiated up to $t$, i.e., $\mathcal N(t) = 0$. Also, we have not included the term $\prod_{j=1}^{J'} p\{N(t_j) \mid \mathcal Y_i(t_{j}), \mathcal N(t_{j-1}), \mathcal X\}$ because, as we will explain in Section~\ref{s:Modeling_est}, this can be ignored under our modeling approach. Even though this type of marginal effect is often used in the context of causal inference, it may be of less interest to urologists because they would typically decide to initiate salvage therapy for patients with elevated PSA. A more clinically interesting setting is $\mathcal Y_i(t)$, i.e., to condition on the longitudinal history of a specific subject $i$:
\begin{eqnarray}
\nonumber \lefteqn{\mbox{ST}^C(t + \Delta t, t) = }\\
\nonumber && \Pr \{T_i^{m(1)} \leq t + \Delta t \mid T_i^m > t, T_i^d > t, \mathcal Y_i(t), \mathcal X_i\}\\
&& - \Pr \{T_i^{m(0)} \leq t + \Delta t \mid T_i^m > t, T_i^d > t, \mathcal Y_i(t), \mathcal X_i\}. \label{Eq:cST_def}
\end{eqnarray}
This effect is \emph{conditional/individualized} in the sense that it is for the group of patients with baseline variables $\mathcal X_i$ and longitudinal history $\mathcal Y_i(t)$, i.e., the same PSA values as for the patient the doctor wants to decide on initiating salvage therapy. Following the discussion in \citet{taylor.et.al:14}, $\mbox{ST}^C(t + \Delta t, t)$ is a conditional causal effect relevant for subject-specific treatment decisions. The difference between the two salvage therapy effects presented above is a bias versus variance trade-off. In particular, the marginal effect (\ref{Eq:mST_def}) marginalizes over a bigger group of patients and will have a smaller variance than the other effects; however, as explained above, it will also be less relevant for the practicing urologist. The conditional effect (\ref{Eq:cST_def}) is the one most relevant to the doctor, but it will have a larger variance because it is based on very little data. A compromise between (\ref{Eq:mST_def}) and (\ref{Eq:cST_def}) is to quantify the salvage therapy effect for patients who had PSA levels above a threshold value $c$ at their last visit, i.e., $\mathcal Y^*(t) = \{Y(t): Y(t) > c\}$:
\begin{eqnarray}
\nonumber \lefteqn{\mbox{ST}^{MC}(t + \Delta t, t) = }\\
\nonumber && \int \biggl [ \Pr \{T^{m(1)} \leq t + \Delta t \mid T^m > t, T^d > t, \mathcal Y^*(t)\}\\
\nonumber && - \Pr \{T^{m(0)} \leq t + \Delta t \mid T^m > t, T^d > t, \mathcal Y^*(t)\}\biggr ]\\ && \prod \limits_{j = 1}^{J'} \dd \tilde G\{Y(t_j), \mathcal X \mid \mathcal Y(t_{j-1})\} \label{Eq:mcST_def}
\end{eqnarray}
where $\tilde G(\cdot)$ denotes the cumulative distribution function for the truncated distribution of $Y$ and $\mathcal X$ satisfying the condition $\mathcal Y^*(t)$. This effect marginalizes over all possible longitudinal histories of PSA that end up having PSA greater than $c$ at time $t$. The effect (\ref{Eq:mcST_def}) offers a compromise by considering a more relevant group of patients, but smaller than the group considered in (\ref{Eq:mST_def}) (i.e., it will have a bigger variance than (\ref{Eq:mST_def})). We should note that by changing the specification of $\mathcal Y^*(t)$, alternative ST effects may be defined. For example, we could define the ST effect for patients with elevated PSA levels in the last $K$ months (e.g., $K = 6$).


\subsection{Assumptions} \label{s:STeffects_ass}
Because of the observational nature of the University of Michigan Prostatectomy Data, we will need a set of assumptions to identify and unbiasedly estimate the salvage therapy effects defined in the previous section. In particular, we will make the standard assumptions for causal inference with observational data, time-varying confounding, and intermediate variables.
\begin{itemize} \itemsep=12pt
\item[-] \emph{Consistency:} The observed outcomes equal the counterfactual outcomes for the actually assigned treatment.

\item[-] \emph{Sequential Exchangeability:} The counterfactual outcomes are independent of the assigned treatment conditionally on the history of PSA measurements, the history of salvage therapy treatments, and baseline covariates, i.e.,
    \[
    T^{m(a)} \perp \!\!\! \perp N(t) \mid \mathcal Y(t), \mathcal N(t), \mathcal X
    \]
    and
    \[
    \mathcal F^{(a)}(v) \perp \!\!\! \perp N(t) \mid \mathcal Y(t), \mathcal N(t), \mathcal X, \;\;\;  v > t,
    \]
    where $a = \{0, 1\}$. Note that we assume here that the decision to start salvage therapy at time $t$ can only causally affect future PSA measurements $Y(v)$, with $v > t$ but not $Y(t)$.
\end{itemize}
The different salvage therapies are sufficiently well-defined in our motivating dataset not to cause any issues with the consistency assumption. The sequential exchangeability assumption also seems to be satisfied in our setting because, as mentioned earlier, urologists typically decide to initiate salvage therapy based on the history of PSA values and possibly other factors such as age and comorbidities (captured in $\mathcal X$). Because of the parametric nature of our modeling approach (presented in the following section), the positivity assumption is not required to identify and estimate the three causal effects we introduced above. However, for the marginal effect (\ref{Eq:mST_def}), we extrapolate beyond the support of the data because the urologists will not prescribe salvage medication for patients with a zero PSA value. Nonetheless, because this effect is of little practical interest, we are not concerned about this extrapolation.


\section{Modeling} \label{s:Modeling}
\subsection{Sub-models Specification}
\label{s:Modeling_def}
We will use the framework of joint models to associate the risk of metastasis with the longitudinal PSA measurements and account for the risk of death. Joint models will account for the endogenous nature of the PSA process, and under the set of assumptions presented in Section~\ref{s:STeffects_ass} will provide valid predictions of the risk of metastasis. We will use the subscript $i$ ($i = 1, \ldots, n$) to denote the subject in all the random variables introduced above. Also, we will use $T_i = \min(T_i^m, T_i^d, C_i)$ to denote the observed event times with $C_i$ denoting the censoring time, and $\delta_i \in \{ 0, 1, 2\}$, where `0' is for censoring, `1' for metastasis, and `2' for death. The deaths we consider as competing events are the ones that occurred before metastasis; deaths and PSA measurements after metastasis are ignored. As explained in the introduction, we expect a drop in PSA levels after the initiation of salvage therapy. Following previous literature in modeling PSA profiles for prostate cancer patients, we specify a linear mixed-effects model for the logarithm of PSA with a change point in the subject-specific trajectories after the initiation of salvage therapy:
\begin{eqnarray}
y_i(t) = \left \{
\begin{array}{ll}
\eta_i(t) + \varepsilon_i(t) = \bx_i(t)^\top \bfbeta + \bz_i(t)^\top \bb_i + \varepsilon_i(t), & t < S_i,\\&\\
\tilde \eta_i(t) + \varepsilon_i(t)\\
 = \eta_i(t) + \Bigl \{ \tilde{\bx}_i(\tilde t)^\top \tilde{\bfbeta} + \tilde{\bz}_i(\tilde t)^\top \tilde{\bb}_i \Bigr \} + \varepsilon_i(t), & t \geq S_i,\\
\end{array}
\right. \label{Eq:PSA_model}
\end{eqnarray}
where the design vectors $\bx_i$ and $\bz_i$ for the fixed $\bfbeta$ and random effects $\bb_i$, respectively describe the subject-specific PSA evolutions before salvage therapy. Analogously, the design vectors $\tilde{\bx}_i$ and $\tilde{\bz}_i$ for the fixed $\tilde{\bfbeta}$ and random effects $\tilde{\bb}_i$, respectively describe the change in the subject-specific PSA evolutions after salvage therapy. The latter design vectors use the relative time variable $\tilde t = t - S_i$. The covariates present in the four design vectors defined above are assumed to be a subset of $\mathcal X_i$. The random effects is the model component that allows for different PSA profiles before and after salvage per patient. The distributional assumptions for the random effects and the error terms are: $\bu_i = (\bb_i^\top, \tilde{\bb}_i^\top)^\top \sim \mathcal N(\mathbf{0}, \bfOmega)$, $\varepsilon_i(t) \sim \mathcal N(0, \sigma^2)$, and $\mbox{cov}\{\bu_i, \varepsilon_i(t)\} = 0$. The variance-covariance matrix $\bfOmega$ is assumed completely unstructured.

Metastasis and death are considered competing risks, and we specify a different hazard model for each one. For metastasis, we postulate the model:
\begin{eqnarray}
\nonumber h_i^m(t) & = & \lim_{\epsilon \rightarrow 0} \epsilon^{-1} \Pr \{t \leq T_i^m < t + \epsilon \mid T_i^m > t, T_i^d > t, \mathcal H_i(t), \mathcal N_i(t), \mathcal X_i \}\\
\nonumber & = & h_0^m(t) \exp \Bigl (\bfpsi_m^\top \bw_i + \gamma_m N_i(t) + \bfalpha_m^\top \bigl [\{1 - N_i(t)\} \times f\{\mathcal H_i(t)\} \bigr ]\\
&& \hspace{3cm} + \; \bfxi_m^\top \bigl [N_i(t) \times g \{ \mathcal H_i(t)\} \bigr ] \Bigr),
\label{Eq:hazard_model}
\end{eqnarray}
where $h_i^m(t)$ is the metastasis-specific hazard function for patient $i$, and $h_0^m(t)$ the baseline hazard. In our model, the logarithm of $h_0^m(t)$ is modeled using penalized B-splines, i.e.,
\[
\log h_0^m(t) = \psi_{h_m,0} + \sum \limits_{q = 1}^Q \psi_{h_m,q} B_q(t),
\]
where $B_q(t)$ denotes the $q$-th basis function of a B-spline with knots $v_1, \ldots, v_{Q+1}$ and $\bfpsi_{h_m}$ the vector of spline coefficients. The design vector $\bw_i$ with the corresponding coefficients vector $\bfpsi$ is for the baseline covariates (subset of $\mathcal X_i$) relevant to the risk of metastasis. The term $\mathcal H_i(t) = \{\eta_i(s); 0 \leq s < \min(S_i, t)\} \bigcup \{\tilde \eta_i(s); S_i \leq s < t\}$ denotes the history of the subject-specific linear predictor. The vector functions $f(\cdot)$ and $g(\cdot)$ determine the functional form for the dependence of the hazard on the PSA evolutions before and after ST, e.g., $f\{\eta_i(t)\} = [\eta_i(t), d \eta_i(t) / dt]^\top$ or $f\{\eta_i(t)\} = [\eta_i(t), \eta_i(t) - \eta_i(t - 1)]^\top$. From this model, we can compute the hazard ratio for salvage therapy for subject $i$, i.e., \begin{equation}
\mbox{HR}(t) = \exp \Bigl (\gamma_m + \bfxi_m^\top  g\{\tilde\eta_i(t)\} - \bfalpha_m^\top f\{\eta_i(t)\} \Bigr ), \quad t > S_i, \label{Eq:HR_ST}
\end{equation}
where in this expression $\eta_i(t)$ represents the expected value of $y_i(t)$ as if subject $i$ had not received salvage therapy at time $S_i$. We note that the metastasis model (\ref{Eq:hazard_model}) could be extended in several ways, including to include interaction terms between $\bw_i$ and $N_i(t)$ or by allowing $\bfpsi_m$ to depend on $t$.

We presume that the PSA is not associated with the hazard of death:
\begin{eqnarray*}
h_i^d(t) & = & \lim_{\epsilon \rightarrow 0} \epsilon^{-1} \Pr \{t \leq T_i^d < t + \epsilon \mid T_i^m > t, T_i^d > t, \mathcal N_i(t), \mathcal X_i\}\\
& = & h_0^d(t) \exp \bigl \{ \bfpsi_d^\top \bw_i + \gamma_d N_i(t) \bigr \},
\end{eqnarray*}
where $h_i^d(t)$ is the death-specific hazard function for patient $i$, and $h_0^d(t)$ is the baseline hazard that is again modeled using penalized B-splines with an associated spline coefficients vector $\bfpsi_{h_d}$. Likewise, interaction terms between $\bw_i$ and $N_i(t)$ could also be included.


\subsection{Estimation} \label{s:Modeling_est}
The longitudinal and event time processes are linked via the random effects to define their joint distribution. We will use a Bayesian approach to fit the postulated joint model. Inference proceeds via the posterior distribution of the parameters $\{\bu_i, \bftheta; i = 1, \ldots, n\}$ given the data $\mathcal D = \{T_i, \delta_i, \bY_i; i = 1, \ldots, n\}$, where $\bftheta$ denotes the vector of model parameters. We start by formulating the joint posterior of $\{\bu_i, \bftheta; i = 1, \ldots, n\}$ and $\bftheta_N$, with $\bftheta_N$ denoting the parameter vector for distribution of the salvage therapy assignment process. Using telescoping, we get:
\begin{eqnarray*}
\lefteqn{p(\bftheta, \bu, \bftheta_N \mid \bT, \bfdelta, \bY, \bN) \propto }\\
&& \prod \limits_{i = 1}^{n} \prod \limits_{j = 1}^{n_i} p\{Y_i(t_{ij}), T_i, \delta_i \mid \mathcal Y_i(t_{i,j-1}), \mathcal N_i(t_{i,j-1}), \mathcal X_i, \bftheta, \bu_i\}\\
&& \times \prod \limits_{j = 1}^{n_i} p\{N_i(t_{ij}) \mid \mathcal Y_i(t_{i,j-1}), \mathcal N_i(t_{i,j-1}), Y_i(t_{ij}), T_i, \delta_i, \mathcal X_i, \bftheta_N, \bu_i\}\\
&& \times \; p(\bu_i \mid \bftheta) \times p(\bftheta) \times p(\bftheta_N)
\end{eqnarray*}
where $\bT$, $\bfdelta$, $\bY$, and $\bN$ denote the vectors of the event times, event indicators, longitudinal measurements, and treatments decisions, respectively, $\mathcal Y_i(t_{i,0}) = \mathcal N_i(t_{i,0}) = \mbox{\O}$, and $p(\cdot)$ denotes an appropriate probability density or probability mass function. Under sequential exchangeability, we have that
\begin{eqnarray*}
\lefteqn{p\{N_i(t_{ij}) \mid \mathcal Y_i(t_{ij}), \mathcal N_i(t_{i,j}), \mathcal F_i^{(a)}(v_{ij}, t_{ij}), T_i^{(a)}, \delta_i^{(a)}, \mathcal X_i, \bftheta_N, \bu_i \} =}\\
&& \hspace{2cm} p\{N_i(t_{ij}) \mid \mathcal Y_i(t_{ij}), \mathcal N_i(t_{i,j-1}), \mathcal X_i, \bftheta_N\},
\end{eqnarray*}
where $\mathcal F_i^{(a)}(v_{ij}, t_{ij})$ denotes the future counterfactual PSA measurements for $v_{ij} > t_{ij}$ and $\{T_i^{(a)}, \delta_i^{(a)}\}$ the counterfactual event times. Hence, assuming that the parameters $\{\bu_i, \bftheta; i = 1, \ldots, n\}$ and $\bftheta_N$ are functionally independent, inference for the posterior distribution of the counterfactual outcomes $\{\bftheta, \bu \mid \bT^{(a)}, \bfdelta^{(a)}, \bY^{(a)}, \bN^{(a)}\}$ can be obtained using the first term (i.e., the observed data model) and ignore the second term. In particular, we get
\begin{eqnarray*}
\lefteqn{p(\bftheta, \bu \mid \bT, \bfdelta, \bY, \bN) \propto}\\
&& \prod \limits_{i = 1}^n \biggl \{ \prod_j \frac{1}{\sqrt{2 \pi \sigma^2}} \exp \Bigl (-\frac{1}{2\sigma^2} \bigl \{y_{ij} - \mu_{ij}(\bftheta, \bu_i, \mathcal N_i)\bigr\}^2 \Bigr) \biggr \}\\
&& \times \mbox{det}(2\pi \bfOmega)^{-1/2} \exp \Bigl(-\frac{1}{2} \bu_i^\top \bfOmega^{-1} \bu_i \Bigr)\\
&& \times \{h_i^{m}(T_i; \bftheta, \bu_i, \mathcal N_i)\}^{I(\delta_i = 1)} \;\; \{h_i^{d}(T_i; \bftheta, \mathcal N_i)\}^{I(\delta_i = 2)}\\
&& \times \exp \Bigl (- \displaystyle \int_0^{T_i}\bigl \{ h_i^{m}(s; \bftheta, \bu_i, \mathcal N_i) + h_i^{d}(s; \bftheta, \mathcal N_i) \bigr \} \; \dd s\Bigr )\\
&& \times \; p(\bftheta),
\end{eqnarray*}
where $\mu_{ij}(\bftheta, \bu_i, \mathcal N_i)$ denotes the mean of the linear mixed model (\ref{Eq:PSA_model}), and $\mbox{det}(A)$ is the determinant of matrix $A$. We use standard priors for $\bftheta$, i.e., normal priors for all regression coefficients $(\bfbeta, \tilde{\bfbeta}, \bfpsi_{h_m}, \bfpsi_m, \gamma_m, \bfalpha_m, \bfxi_m, \bfpsi_{h_d}, \bfpsi_d, \gamma_d)$, inverse-Gamma priors for $\sigma^2$ and the diagonal elements of $\bfOmega$, and the LKJ prior for the correlation matrix of the random effects \citep{lewandowski.et.al:09}. To ensure smoothness of the baseline hazard functions $h_0^m(t)$ and $h_0^d(t)$, we postulate a `penalized' prior distribution for the regression coefficients $\bfpsi_{h_m}$ and $\bfpsi_{h_d}$ (we only show the formulation for the former):
\[
p(\bfpsi_{h_m} \mid \tau_m) \propto \tau_m^{\rho(K)/2}\exp \Bigl (-\frac{\tau_m}{2}
\bfpsi_{h_m}^\top \bK \bfpsi_{h_m} \Bigr ),
\]
where $\tau_m$ is the smoothing parameter that takes a $\mbox{Gamma}(5, 0.05)$ hyper-prior in order to ensure a proper posterior for $\bfpsi_{h_m}$, $\bK = \Delta_r^\top \Delta_r$, where $\Delta_r$ denotes $r$-th difference penalty matrix, and $\rho(\bK)$ denotes the rank of $\bK$.

We use a Markov chain Monte Carlo (MCMC) approach to obtain samples from the posterior distribution for all model parameters and the random effects. This algorithm is implemented in the \textsf{R} package \textbf{JMbayes2} \citep{JMbayes2} that we used to fit the model and calculate the salvage therapy effects.


\section{Salvage Therapy Effects Estimation} \label{s:STeffects_Est}
\subsection{Estimates} \label{s:STeffects_Est_estimates}
We start with the estimation of the conditional effect (\ref{Eq:cST_def}) that we calculate for a specific patient with longitudinal history $\mathcal Y_i(t)$ and covariates $\mathcal X_i$. Both terms in the definition of this effect are posterior predictive distributions, which under the postulated joint model are written as
\begin{eqnarray}
\nonumber \lefteqn{\Pr \{T_i^{(a)} \leq t + \Delta t, \delta_i^{(a)} = 1 \mid T_i > t, \mathcal Y_i(t), \mathcal X_i\} = }\\
\nonumber && \displaystyle \int \displaystyle \int \Pr \{T_i^{(a)} \leq t + \Delta t, \delta_i^{(a)} = 1 \mid T_i > t, \bu_i, \mathcal X_i, \bftheta\}\\
&& \hspace{2cm}\times \; p\{\bu_i \mid T_i > t, \mathcal Y_i(t), \mathcal X_i, \bftheta\} \; p(\bftheta \mid \mathcal D) \; \dd\bu_i \dd\bftheta, \label{Eq:cST_est}
\end{eqnarray}
where $a = \{0, 1\}$, and the term $p(\bftheta \mid \mathcal D)$ denotes the posterior distribution for the model parameters $\bftheta$. Again for simplicity of exposition, we have omitted the conditioning on $\mathcal N(t) = 0$. We should note that these risk predictions are marginalized over both the parameters and the random effects. The first term in the integrand is written as:
\[
\Pr \{T_i^{(a)} \leq t + \Delta t, \delta_i^{(a)} = 1 \mid T_i > t, \bu_i, \mathcal X_i, \bftheta\} = \mathcal A / \mathcal B,
\]
where
\begin{eqnarray*}
\lefteqn{\mathcal A = \displaystyle \int_t^{t + \Delta t} h_i^{m(a)}(v) \exp \Bigl (- \displaystyle \int_t^v \bigl \{ h_i^{m(a)}(s) + h_i^{d(a)}(s) \bigr \} \; \dd s}\\
&& \quad - \displaystyle \int_0^t \bigl \{ h_i^{m(0)}(s) + h_i^{d(0)}(s) \bigr \} \; \dd s \Bigr ) \; \dd v,
\end{eqnarray*}
and
\[
\mathcal B = \displaystyle \exp \Bigl (- \displaystyle \int_0^t \bigl \{ h_i^{m(0)}(s) + h_i^{d(0)}(s) \bigr \} \; \dd s\Bigr ),
\]
with $h_i^{m(1)}(t) = h_0^m(t) \exp (\bfpsi_m^\top \bw_i + \gamma_m + \bfxi_m^\top  g\{\tilde\eta_i(t)\})$ and $h_i^{m(0)}(t) = h_0^m(t) \exp(\bfpsi_m^\top \bw_i + \bfalpha_m^\top f\{\eta_i(t)\})$.
The counterfactual hazard functions for death $h_i^{d(1)}(t)$ and $h_i^{d(0)}(t)$ are defined analogously. In particular, $h_i^{d(1)}(t) = h_0^d(t) \exp (\bfpsi_d^\top \bw_i + \gamma_d)$ and
$h_i^{d(0)}(t) = h_0^d(t) \exp (\bfpsi_d^\top \bw_i)$. In the specification of the conditional distribution of the random effects given the observed information $\{T_i > t, \mathcal Y_i(t), \mathcal X_i\}$, it is assumed that salvage therapy has not been initiated by time $t$. Combining the integral equations presented above, we can estimate (\ref{Eq:cST_est}) using the following Monte-Carlo scheme:
\begin{enumerate}
\item[S1] We sample $\breve{\bftheta}^{(l)}$ from the MCMC sample of the posterior distribution $[\bftheta \mid \mathcal D]$.

\item[S2] We sample $\breve{\bu}_i^{(l)}$ from the posterior distribution $[\bu_i \mid T_i > t, \mathcal Y_i(t), \mathcal X_i, \breve{\bftheta}^{(l)}]$. Because this distribution cannot be written in closed-form, we sample from it using the Metropolis-Hastings algorithm.

\item[S3] We calculate the term $\pi_i^{(l)} (t + \Delta t \mid t, a) = \Pr \{T_i^{(a)} \leq t + \Delta t, \delta_i^{(a)} = 1 \mid T_i > t, \breve{\bu}_i^{(l)}, \mathcal X_i, \breve{\bftheta}^{(l)}\}$. The integrals in the definition of the overall survival functions are approximated numerically using the 15-point Gauss-Kronrod quadrature rule.
\end{enumerate}
We repeat Steps S1--S3, $L$ times, and we take as an estimate of $\mbox{ST}_i^C(t + \Delta t, t)$, the mean over the Monte-Carlo samples, i.e.,
\begin{equation}
\widehat{\mbox{ST}}_i^C(t + \Delta t, t) = \frac{1}{L} \sum_{l = 1}^L \pi_i^{(l)} (t + \Delta t \mid t, a = 1) - \pi_i^{(l)} (t + \Delta t \mid t, a = 0). \label{Eq:cST_MCest}
\end{equation}

The estimation of the marginal effects (\ref{Eq:mST_def}) and (\ref{Eq:mcST_def}) proceeds by averaging the conditional effects over the respective groups of patients in the sample. In particular, for (\ref{Eq:mST_def}), we define $\mathcal R(t)$ to denote the subset of patients at risk at time $t$. For each patient in $\mathcal R(t)$, we calculate $\widehat{\mbox{ST}}_i^C(t + \Delta t, t)$. Then, we obtain the estimator
\[
\widehat{\mbox{ST}}^M(t + \Delta t, t) = n_r^{-1} \sum_{i: i \in R(t)} \widehat{\mbox{ST}}_i^C(t + \Delta t, t),
\]
where $n_r$ denotes the number of subjects in $\mathcal R(t)$. For (\ref{Eq:mcST_def}), the summation is over $\mathcal R^*(t)$ that denotes the subset of patients at risk at time $t$ and had longitudinal histories that satisfy the definition of $\mathcal Y^*(t)$ ($n_{r^*}$ is defined analogously).


\subsection{Variance} \label{s:STeffects_Est_variance}
To derive the variance of $\widehat{\mbox{ST}}^M(t + \Delta t, t)$ and $\widehat{\mbox{ST}}^{MC}(t + \Delta t, t)$, we need to take into account that they are a function of both the parameters $\bftheta$ and the data $\mathcal D$ (in the definition of $\mathcal D$ here we also include $\{\mathcal X_i, i = 1, \ldots, n_r\}$). The variance of both estimators can be derived using similar arguments, and here we will only show how to calculate the variance of the former. To make the dependence on the data $\mathcal D$ more explicit, we write the estimator of the marginal salvage therapy effect as:
\[
\widehat{\mbox{ST}}^M \bigl ( t + \Delta t, t; \mathcal D \bigr ) = E_{\bftheta \mid \mathcal D} \Bigl \{ \mbox{ST}^M \bigl ( t + \Delta t, t; \bftheta, \mathcal D \bigr ) \Bigr \}.
\]
To account for both the variability in the parameters and the sampling variability, our target variance is
\begin{eqnarray}
\nonumber \lefteqn{\mbox{var}_{\mathcal D} \bigl \{ \widehat{\mbox{ST}}^M \bigl ( t + \Delta t, t; \bftheta, \mathcal D \bigr ) \bigr \}}\\
&& = \mbox{var}_{\mathcal D} \biggl [ E_{\bftheta \mid \mathcal D} \Bigl \{ \mbox{ST}^M \bigl (t + \Delta t, t; \bftheta, \mathcal D \bigr) \Bigr \} \biggr ]. \label{Eq:var_mST}
\end{eqnarray}
We will adapt the approach of \citet{antonelli.et.al:20} to obtain an estimate of (\ref{Eq:var_mST}). More specifically, we let $\{\mathcal D^{(1)}, \ldots, \mathcal D^{(M)}\}$ to denote $M$ datasets sampled with replacement from $\mathcal D$. For each of these datasets, we calculate
\begin{eqnarray*}
\lefteqn{\widehat{\mbox{ST}}^M \bigl ( t + \Delta t, t; \bftheta, \mathcal D^{(m)} \bigr ) =}\\
&& E_{\bftheta \mid \mathcal D} \Bigl \{ \mbox{ST}^M \bigl (t + \Delta t, t; \bftheta, \mathcal D^{(m)} \bigr ) \Bigr \}, m = 1, \ldots, M.
\end{eqnarray*}
Note that the expectation is taken with respect to the posterior distribution of the parameters using the original data $\mathcal D$ (i.e., we do not refit the model for each dataset $\mathcal D^{(m)}$). Calculating the sample variance of these $M$ estimates we obtain
\begin{equation}
\mbox{var}_{\mathcal D^{(m)}} \biggl [ E_{\bftheta \mid \mathcal D} \Bigl \{ \mbox{ST}^M \bigl (t + \Delta t, t; \bftheta, \mathcal D^{(m)} \bigr) \Bigr \} \biggr ]. \label{Eq:var_mST_est_naive}
\end{equation}
Even though (\ref{Eq:var_mST_est_naive}) resembles our target variance, it ignores the variability in the posterior due to the different samples $\{\mathcal D^{(1)}, \ldots, \mathcal D^{(M)}\}$. Hence, to get correct inferences, we use the correction term:
\begin{eqnarray}
\nonumber \lefteqn{\widehat{\mbox{var}}_{\mathcal D} \biggl [ E_{\bftheta \mid \mathcal D} \Bigl \{ \mbox{ST}^M \bigl (t + \Delta t, t; \bftheta, \mathcal D \bigr) \Bigr \} \biggr ] = }\\
\nonumber && \mbox{var}_{\mathcal D^{(m)}} \biggl [ E_{\bftheta \mid \mathcal D} \Bigl \{ \mbox{ST}^M \bigl (t + \Delta t, t; \bftheta, \mathcal D^{(m)} \bigr) \Bigr \} \biggr ]\\
&& + \mbox{var}_{\bftheta \mid \mathcal D} \Bigl \{ \mbox{ST}^M \bigl (t + \Delta t, t; \bftheta, \mathcal D \bigr) \Bigr \}. \label{Eq:var_mST_est}
\end{eqnarray}

To derive the variance of $\widehat{\mbox{ST}}_i^C(t + \Delta t, t)$ we need to account for the sampling variability in $\mathcal Y_i(t)$, i.e., the PSA values patient $i$ would have shown if we `cloned' him. To account for this variability, we cannot use the same idea as in $\widehat{\mbox{ST}}^M(t + \Delta t, t)$ above because we cannot obtain samples with replacement from $\mathcal Y_i(t)$. As an alternative, we employ a parametric Bootstrap approach, i.e., we create different versions of the history $\{\mathcal Y_i^{(1)}(t), \ldots, \mathcal Y_i^{(M)}(t)\}$ using the Monte-Carlo scheme:
\begin{enumerate}
\item[Q1] We sample $\ddot{\bftheta}^{(l)}$ from the MCMC sample of the posterior distribution $[\bftheta \mid \mathcal D]$.

\item[Q2] We sample $\ddot{\bu}_i^{(l)}$ from the posterior distribution $[\bu_i \mid T_i > t, \mathcal Y_i(t), \mathcal X_i, \ddot{\bftheta}^{(l)}]$.

\item[Q3] We simulate $\mathcal Y_i^{(m)}(t)$ using independent draws from $[y_i \mid \ddot{\bu}_i, \mathcal X_i, \ddot{\bftheta}^{(l)}]$, that is from the linear mixed model (\ref{Eq:PSA_model}) with $t < S_i$.
\end{enumerate}
Subsequently, we use the same procedure as for the variance of $\widehat{\mbox{ST}}^M( t + \Delta t, t; \mathcal D)$. Namely, along the lines of (\ref{Eq:var_mST_est}) we use:
\begin{eqnarray*}
\lefteqn{\widehat{\mbox{var}}_{\mathcal Y_i} \biggl [ E_{\bftheta \mid \mathcal Y_i} \Bigl \{ \mbox{ST}^C \bigl (t + \Delta t, t; \bftheta, \mathcal Y_i(t) \bigr) \Bigr \} \biggr ] = }\\
&& \mbox{var}_{\mathcal Y_i^{(m)}} \biggl [ E_{\bftheta \mid \mathcal Y_i} \Bigl \{ \mbox{ST}^C \bigl (t + \Delta t, t; \bftheta, \mathcal Y_i^{(m)}(t) \bigr) \Bigr \} \biggr ]\\
&& + \mbox{var}_{\bftheta \mid \mathcal Y_i} \Bigl \{ \mbox{ST}^C \bigl (t + \Delta t, t; \bftheta, \mathcal Y_i(t) \bigr) \Bigr \}.
\end{eqnarray*}
Again, the first term accounts for the variability in $\mathcal Y_i(t)$ and the second for the parameters' variability. The calculation of the first term entails obtaining the Monte Carlo estimate (\ref{Eq:cST_MCest}) for each realization $\mathcal Y_i^{(m)}(t)$ (i.e., we have nested Monte Carlo schemes). Because of its parametric nature, this estimator relies more heavily on the model's formulation.


\section{University of Michigan Prostatectomy Data - Analysis} \label{s:UMPD_Analysis}
We return to the UMP dataset, which we will use to estimate the salvage therapy effects introduced in Section~\ref{s:STeffects}. More information on the dataset and some pre-processing we applied is given in Web Section~1.1. In the original version of the dataset, some patients received salvage therapy multiple times. For those patients, we have only considered the first time they received salvage therapy and used in the analysis only the PSA measurements taken up to 1.5 years after salvage therapy.

The joint model we fitted to the final dataset had the following specification. For the PSA trajectories, we use a modified version of the generic model (\ref{Eq:PSA_model}), i.e.,
\[
\log \{\var{PSA}(t_{ij}) + 1 \} = \left \{
\begin{array}{l}
\eta_i(t_{ij}) + \varepsilon_i(t_{ij}) = \beta_{0i} + \sum_{k = 1}^8 \beta_{ki} \mathcal{B}_k(t_{ij}, v)\\ + \bx_{base,i}^\top \bflambda + \varepsilon_i(t_{ij}), \; t_{ij} \leq S_i,\\
\tilde{\eta_i(t_{ij})} + \varepsilon_i(t_{ij}) = \eta_i(t_{ij})\\
+ \tilde \beta_{0i} + \tilde \beta_{1i} (t_{ij} - S_i) + \bx_{base,i}^\top \tilde{\bflambda} + \varepsilon_i(t_{ij}), \; t_{ij} > S_i.\\
\end{array}
\right.
\]
The subject-specific coefficients, $\beta_{0i}, \ldots, \beta_{8i}$, $\tilde \beta_{0i}$, and $\tilde \beta_{1i}$ are decomposed into a fixed and random part. The combined random-effects vector $\bb_i = (b_{0i}, \ldots, b_{8i}, \tilde{b}_{0i}, \tilde{b}_{1i})^\top$ is assumed to follow a multivariate normal distribution with mean zero and covariance matrix $\bfOmega$, and is independent of the error terms $\varepsilon_{ij}$ that follow a normal distribution with mean zero and variance $\sigma^2$. In both branches of the model, and via the terms $\bx_{base,i}^\top \bflambda$ and $\bx_{base,i}^\top \tilde{\bflambda}$ we control for the baseline covariates age, baseline PSA, Gleason score, and the Charlson comorbidity index. After prostatectomy, most patients will exhibit close to zero PSA levels; however, for some patients, at some point, PSA will start to rapidly increase, triggering the urologists to initiate salvage therapy. The time PSA will increase will differ per patient, requiring a flexible model to capture these evolutions. Hence, we postulate a nonlinear model with a cubic B-spline for the time variable, with internal knots placed at $v = \{0.5, 1, 3, 5, 7, 9\}$ years after prostatectomy and boundary knots placed at 0 and 17.1 years after prostatectomy (i.e., the terms $\mathcal{B}_k(t_{ij}, v)$). After salvage, we assume that the PSA will drop by $\tilde \beta_{0i}$ and then linearly increase before metastasis. We allow for a nonlinear PSA profile during follow-up with a change in the linear slope after salvage because of the limited available information in the data for $t > S_i$. Before salvage, patients had a median of six PSA measurements (IQR = 6), whereas, after salvage, the median of available PSA measurements was three (IQR = 0).

For the hazard of metastasis, we postulate the relative risk models:
\[
h_i^m(t) = h_0^m(t) \exp \bigl \{ \bfpsi_m^\top \bw_{mi} + \mathcal{A}(t) \bigr\},
\]
where for the time-varying component $\mathcal{A}(t)$ we consider the following versions
\begin{eqnarray*}
M_1: \mathcal{A}(t) & = & \left \{
\begin{array}{l}
\alpha_{m1} \eta_i(t), \quad t \leq S_i,\\
\gamma_{m1} (t - S_i) + \gamma_{m2} \{(t - S_i) \times \var{basePSA}_i\}\\ \quad + \xi_{m1} \tilde{\eta}_i(t), \quad t > S_i;\\
\end{array}
\right.\\\\
M_2: \mathcal{A}(t) & = & \left \{
\begin{array}{ll}
\alpha_{m1} \eta_i(t) + \alpha_{m2} \displaystyle \frac{\dd \eta_i(t)}{\dd t}, \quad t \leq S_i,\\
\gamma_{m1} (t - S_i) + \gamma_{m2} \{(t - S_i) \times \var{basePSA}_i\}\\ + \xi_{m1} \tilde{\eta}_i(t), \quad t > S_i;\\
\end{array}
\right.\\\\
M_3: \mathcal{A}(t) & = & \left \{
\begin{array}{ll}
\alpha_{m1} \eta_i(t) + \alpha_{m2} \displaystyle \frac{\dd \eta_i(t)}{\dd t}, \quad t \leq S_i,\\
\gamma_{m1} (t - S_i) + \gamma_{m2} \{(t - S_i) \times \var{basePSA}_i\}\\
+ \xi_{m2} \tilde \beta_{0i} + \xi_{m3} \tilde \beta_{1i}, \quad t > S_i.\\
\end{array}
\right.\\\\
M_4: \mathcal{A}(t) & = & \left \{
\begin{array}{ll}
\alpha_{m1} \eta_i(t) + \alpha_{m3} \{ \int_0^t \eta_i(v) \, dv \big / t\}, \quad t \leq S_i,\\
\gamma_{m1} (t - S_i) + \gamma_{m2} \{(t - S_i) \times \var{basePSA}_i\}\\ + \xi_{m2} \tilde \beta_{0i} + \xi_{m3} \tilde \beta_{1i}, \quad t > S_i.\\
\end{array}
\right.
\end{eqnarray*}
Coefficient $\alpha_{m1}$, $\alpha_{m2}$ and $\alpha_{m3}$ quantify the association between the current value, current slope, and average $\log(\var{PSA} + 1)$ in the period before salvage and the hazard of metastasis, respectively. Coefficient $\xi_{m1}$ quantifies the association of the current value of $\log(\var{PSA} + 1)$ after salvage and the instantaneous risk of metastasis; coefficients $\xi_{m2}$ and $\xi_{m3}$ quantify the association between the drop of PSA just after salvage, and the change in the linear slope of $\log(\var{PSA} + 1)$ after salvage and the instantaneous risk of metastasis, respectively. Also, the coefficient $\gamma_{m1}$ quantifies the association between the length of the period the patient was on salvage therapy and the hazard of metastasis, and coefficient $\gamma_2$ the interaction between the time a patient has been on salvage and baseline PSA. The baseline covariates part $\bfpsi_m^\top \bw_{mi}$ includes the effects of age, baseline PSA, the Gleason score, and the Charlson comorbidity index. The baseline hazard is approximated with B-splines, as explained earlier. For the hazard of death, we assume the model:
\begin{eqnarray*}
h_i^d(t) & = & h_0^d(t) \exp \bigl \{ \psi_{d1} \var{(Age - 50)} + \gamma_{d1} N_i(t) \bigr \}.
\end{eqnarray*}
Samples from the posterior distribution of the model parameters from these four joint models have been obtained from the MCMC algorithm provided in the R package \textbf{JMbayes2} using three parallel chains started from different initial values. The priors used for the various parameters are presented in the supplementary material. Trace-plots and the potential scale reduction factor ($\hat{R}$) showed satisfactory convergence of the MCMC algorithm. In particular, for all model parameters the $\hat{R}$-values were smaller than 1.09.

We first evaluate the model fit before focusing on the results. All fitted joint models showed a similar fit to the longitudinal data; therefore, we only show the results from model $M_1$. Web Figures~3--5 show the observed data and fitted longitudinal trajectories for 26 selected patients from the dataset. These patients have been selected to showcase different profiles seen in the data, including patients who did and did not initiate salvage therapy. We observe from these figures that the postulated longitudinal submodel provides a good fit to the observed data. As previously observed \citep{loic.et.al:19}, this is an important aspect in joint models. In addition, using residuals plots we assessed the normality and homoscedastiscity assumptions for the error terms; these plots did not indicate any violations of these assumptions. Table~\ref{Tab:Info} shows the Deviance Information Criterion (DIC) and the Watanabe-Akaike information criterion (WAIC) for the four fitted joint models.
\begin{table}[ht]
\centering
\caption{Deviance information criterion (DIC), the Watanabe-Akaike information criterion, and the log pseudo marginal likelihood (LPML) for the fitted joint models $M_1 - M_4$.}
\label{Tab:Info}
\begingroup\small
\begin{tabular}{rrrr}
\hline
& DIC & WAIC & LPML \\
\hline
$M_4$ & 23505.49 & 12902244.27 & $-$130702.53 \\
$M_3$ & 25920.73 & 17954385.81 & $-$177836.90 \\
$M_1$ & 43554.78 & 33007334.39 & $-$263741.71 \\
$M_2$ & 91596.92 & 44668135.08 & $-$270099.58 \\
\hline
\end{tabular}
\endgroup
\end{table}
According to both WAIC and DIC, Model $M_4$ provides better predictions compared to the other three models.

For the event process, we show in Web~Table~2 the posterior means and the corresponding 95\% credible intervals for the coefficients of the relative risk models for metastasis and death. All models suggest that before initiating salvage, high PSA levels at time $t$, high PSA velocity at $t$, and high average PSA from baseline to $t$ translate to a greater risk of metastasis at $t$. Models $M_1$ and $M_2$ indicate that after salvage has been initiated, high PSA levels and high PSA velocity at time $t$ translate to a higher risk of metastasis. Models $M_3$ and $M_4$ suggest that the greater the drop in PSA levels after salvage initiation, the lower the risk of metastasis afterward, and the greater the PSA velocity after salvage, the higher the risk of metastasis. For the longitudinal process, the corresponding coefficients are shown in Web Table~3. We observe that the baseline PSA, the Gleason score, and the Charlson comorbidity index seem to be associated with the level of post-surgery PSA. There is no indication that the effect of these confounders change after the initiation of salvage therapy.

\begin{figure}
\centering{\includegraphics[width=\textwidth]{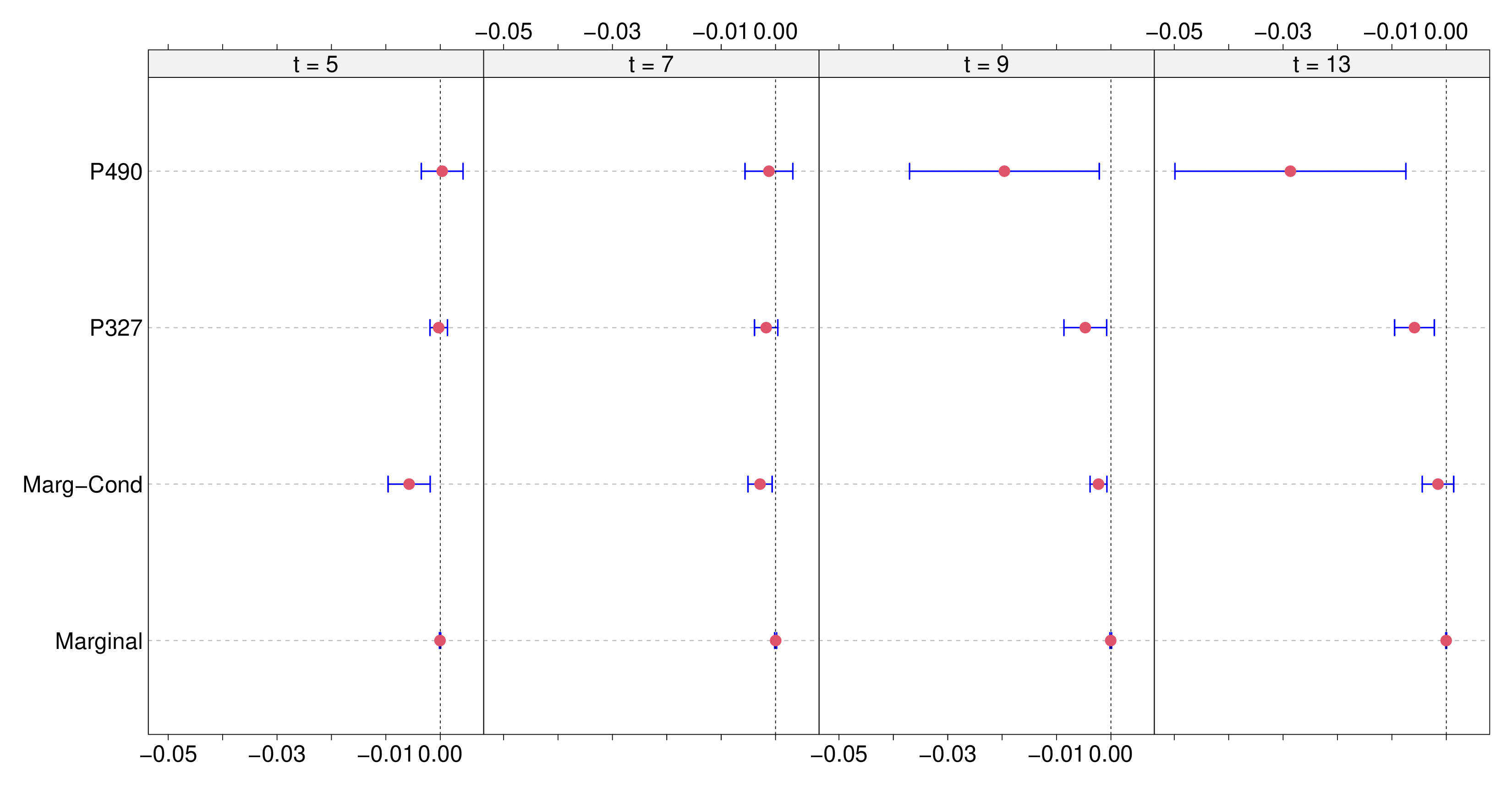}}
\caption{Salvage therapy effects for follow-up times $t = 5, 7, 9$ and $13$ years and $\Delta t = 2$ under model $M_1$. For Patients~490 and 327, conditional causal effects are shown. The marginal-conditional causal effect is for patients who had a PSA value greater or equal to 0.5 ng/mL at their last visit. The marginal effect is for all patients at risk at the corresponding $t$.}
\label{Fig:effects_M1}
\end{figure}

\begin{figure}
\centering{\includegraphics[width=\textwidth]{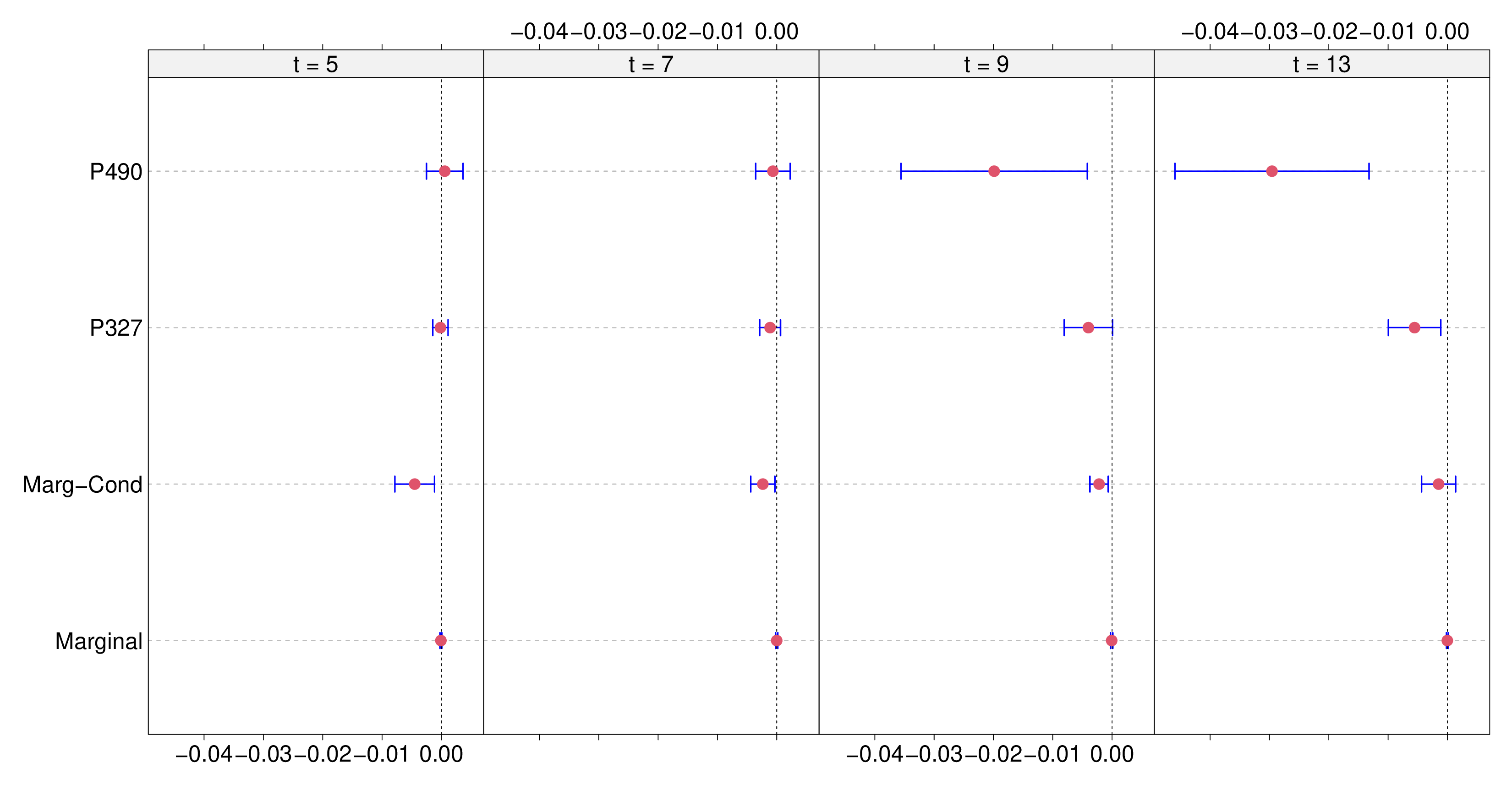}}
\caption{Salvage therapy effects for follow-up times $t = 5, 7, 9$ and $13$ years and $\Delta t = 2$ under model $M_2$. For Patients~490 and 327, conditional causal effects are shown. The marginal-conditional causal effect is for patients who, at their last visit, had a PSA value greater or equal to 0.5 ng/mL. The marginal effect is for all patients at risk at the corresponding $t$.}
\label{Fig:effects_M2}
\end{figure}

\begin{figure}
\centering{\includegraphics[width=\textwidth]{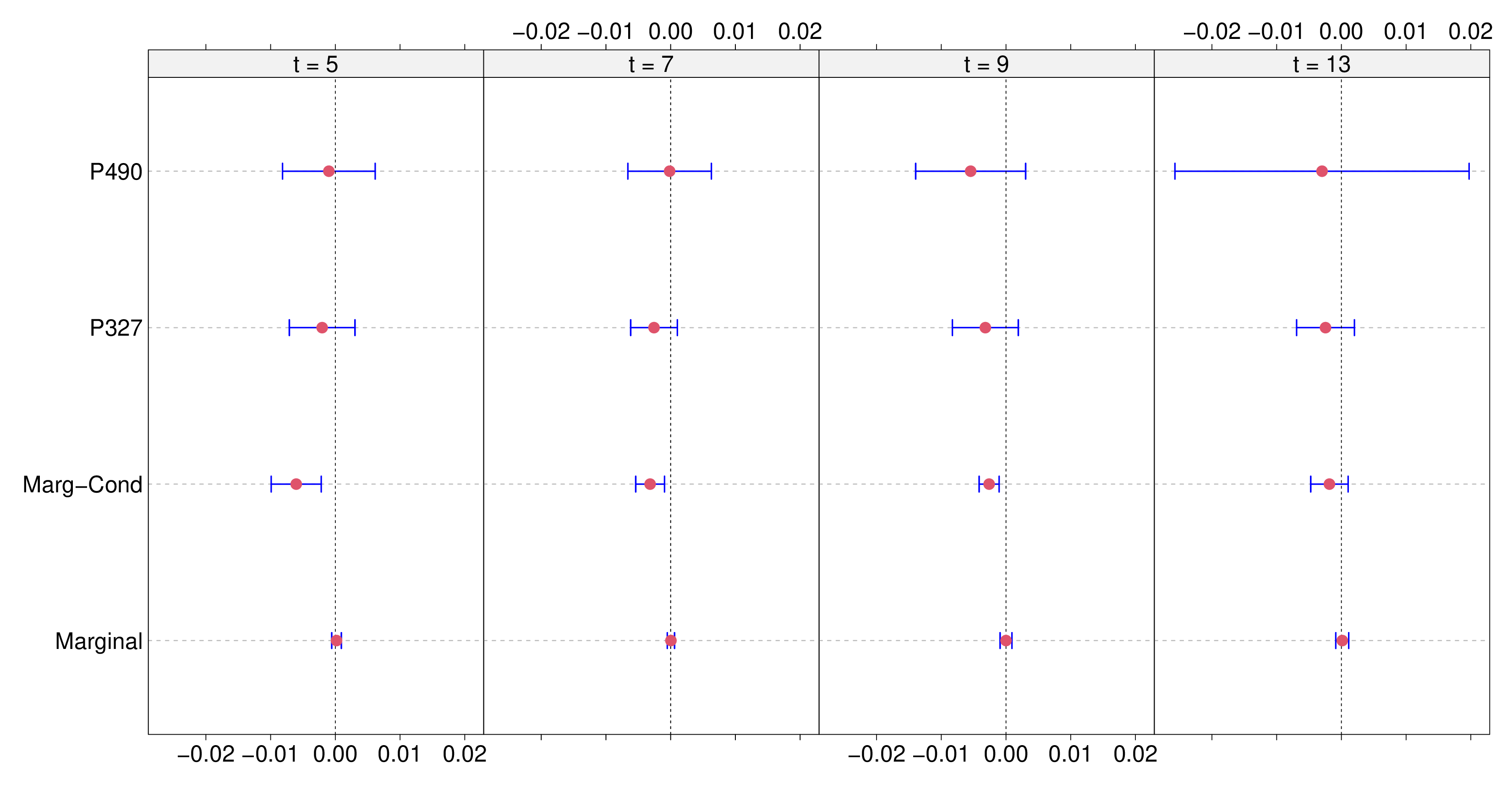}}
\caption{Salvage therapy effects for follow-up times $t = 5, 7, 9$ and $13$ years and $\Delta t = 2$ under model $M_3$. For Patients~490 and 327, conditional causal effects are shown. The marginal-conditional causal effect is for patients who had a PSA value greater or equal to 0.5 ng/mL at their last visit. The marginal effect is for all patients at risk at the corresponding $t$.}
\label{Fig:effects_M3}
\end{figure}

\begin{figure}
\centering{\includegraphics[width=\textwidth]{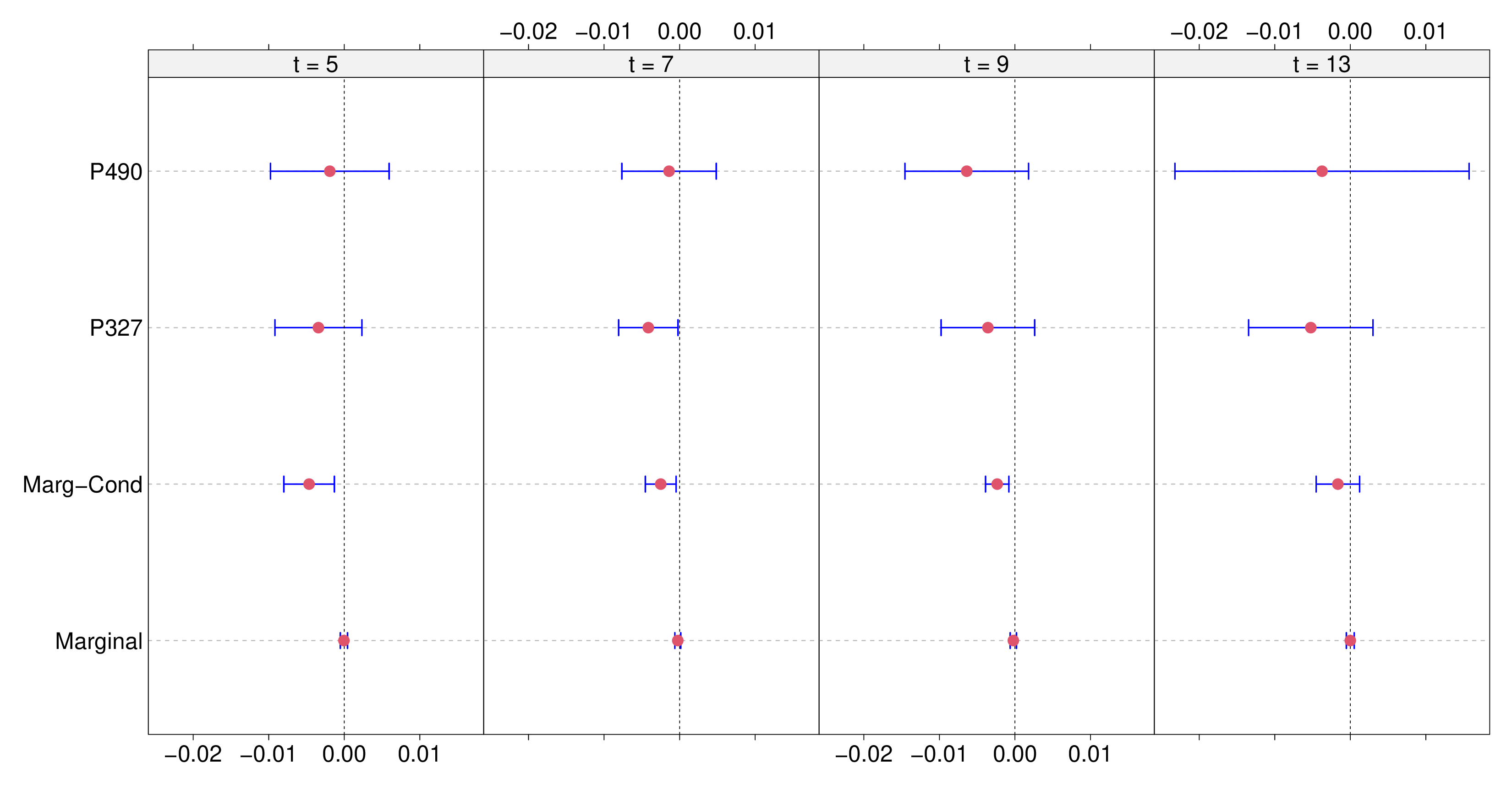}}
\caption{Salvage therapy effects for follow-up times $t = 5, 7, 9$ and $13$ years and $\Delta t = 2$ under model $M_4$. For Patients~490 and 327, conditional causal effects are shown. The marginal-conditional causal effect is for patients who had a PSA value greater or equal to 0.5 ng/mL at their last visit. The marginal effect is for all patients at risk at the corresponding $t$.}
\label{Fig:effects_M4}
\end{figure}

We now turn our focus on estimating causal salvage therapy effects from the fitted joint models. To contrast the different effect types, we show the conditional effect (\ref{Eq:cST_def}) for Patients~490 and 327, the marginal-conditional effect (\ref{Eq:mcST_def}) that averages over the group of patients who at their last visit has a PSA value greater than 0.5 ng/mL, and the marginal effect (\ref{Eq:mST_def}) that averages over all patients irrespective of their PSA values. These effects have been calculated at $t = 5, 7, 9$ and $13$ years after the initial surgery and for a $\Delta t = 2$-year medically relevant time window. The 95\% confidence intervals are calculated using the variance estimate presented in Section~\ref{s:STeffects_Est_variance}. The number of patients based on which the marginal and marginal-conditional effects have been calculated is shown in Web Table~1. Figures~\ref{Fig:effects_M1}--\ref{Fig:effects_M4} present these results.
The estimated values of the conditional and marginal-conditional salvage therapy effects tend to be negative, as expected, since salvage therapy is considered to be beneficial but are also small in magnitude.  They are small because the probability of metastasis in the next two years is itself small, even in the absence of salvage therapy, so the room for improvement is also small. The results for the salvage therapy effects and across models align with the points raised in Section~\ref{s:STeffects_defs}. In particular, we observe that the marginal effects are the smallest in magnitude, followed by the marginal-conditional and conditional effects. The reverse is observed regarding the variance of these effects, with the marginal effects having the smallest variance and the conditional effects the largest. Also, we observe that the conditional effects are more adaptable to the shape of the PSA profile. For example, while the marginal and marginal-conditional effects become smaller in magnitude at later follow-up times, the conditional effect for Patient~490 increases in size because this subject shows a steeply increasing PSA trajectory. This also happens for Patient~327 but to a lesser degree because his PSA profile is less steep. The variance of the conditional effects also increases with increasing PSA values. This reflects the fact that because in the sample, the majority of patients showed stable PSA profiles close to zero, the model is less ``certain'' for the shape of increasing PSA trajectories.

In this section we estimated the marginal-conditional effect at time $t$, by averaging the causal effects of individuals who had their last PSA greater than 0.5. A different marginal-conditional effect would have been obtained if we had used different criteria for who to average over. For example, we could have used a range of PSA, say 0.5 to 4.0. The criteria could have also included requirements such as Gleason score was at least 8 and age was less than 75. We might also exclude from the set any patient whose last PSA was not current, e.g was more than two years ago. Restrictions such as these could make the causal estimates more targeted to an individual patient, but at the expense of greater variance.


\section{Simulation} \label{s:Simulation}
We performed a simulation study to assess the performance of the approaches presented in the previous sections. Note that because the causal effects (\ref{Eq:mST_def})--(\ref{Eq:mcST_def}) entail most of the model parameters in their definition, it is very challenging to simulate data with specific values for the causal effects. This prevents us from setting up a simulation to assess whether specific values for the causal effects can be unbiasedly estimated. However, the key assumption behind our approach is that joint models can be unbiasedly estimated under time-varying confounding and do not require to include a model for the treatment initiation process. This property of joint models has not been established before to our knowledge. Therefore, we focus here on validating the finite sample performance of joint models when the decision to initiate salvage therapy heavily depends on past PSA values. By setting the decision to perform salvage therapy on previous PSA values, we thus, mimic a causal relationship between the decision to perform salvage therapy and the history of PSA. Furthermore, to reduce the computational burden, we opted for a simplified setting with no additional covariates and linear time evolutions. Also, when we simulate PSA values we have not imposed the constraint that they should be positive.

More specifically, we assumed 1000 subjects and then randomly selected follow-up visits, $t_{ij}$ from a uniform distribution between 0 and 20. To mimic the causal relationship between salvage therapy and PSA history, the timing of salvage therapy was assumed to depend on the value of PSA. Specifically, if PSA was smaller than 2 ng/mL, the probability of receiving salvage at that visit time is 0.01; when PSA was in the interval (2 ng/mL, 4 ng/mL), the probability of receiving treatment was 0.5, and for PSA greater than 4 ng/mL, the probability was set to 0.9. If salvage therapy was not given according to the model described previously, the subject was assumed not to undergo salvage therapy but still be at risk for metastasis or death. More details for the settings of the simulation study are given in Web Section~2 in the supplementary material. We then simulated 300 datasets under three different scenarios for the association structure between features of the longitudinal PSA values and the hazard of metastasis. In Scenario~1, we considered an association with the current value of PSA; in Scenario~2, the current value and the current slope before salvage therapy and only the current value after salvage therapy; and in Scenario~3, the value and cumulative effect before salvage therapy  and the current value after salvage therapy. We assumed no association between the PSA history and the hazard of death. The detailed model specification and parameter values are provided in the supplementary material, along with visualizations of the simulated data.

The simulation study results are presented in Web Figures~6--8 and Web Table~2. The results suggest that joint models can unbiasedly estimate the parameters of the joint distribution of the longitudinal and event time outcomes in the presence of time-varying confounding. Hence, we expect that well-specified joint models can be used to estimate causal effects for time-varying treatments.


\section{Discussion}
\label{s:Discuss}
In this paper, we have showcased how causal effects for time-varying treatments (or exposures) can be estimated using the framework of joint models for longitudinal and time-to-event data. These models will account for time-varying confounding without requiring an explicit specification of a model for the probability of receiving treatment conditional on the history of longitudinal confounders and past treatments. The causal effects (\ref{Eq:mST_def})--(\ref{Eq:mcST_def}) are in the flavor of the parametric G-formula and, by conditioning on different specifications of the longitudinal PSA history correspond to different targets of inference.

Our approach relies on parametric assumptions and is expected to use the available data efficiently. However, the estimated causal effects will be biased if these assumptions are seriously violated. To minimize the chance of biased estimated effects, performing a thorough check of the model's assumption using residual plots and evaluating the model's fit with figures such as Web Figures~3--5 is advisable. As we have seen in Section~\ref{s:Modeling_est}, an advantage of our full likelihood approach is that we do not need to model the salvage therapy initiation process. This property of our approach also holds for all other processes that may depend on the observed PSA history. For example, if the treating urologists decide to change the visiting process (i.e., when patients should come back for their next PSA test), and if this decision is solely based on the past observed PSA history of the patient (and possibly covariates), then our modeling approach will still provide valid results without requiring modeling this process. The same consideration also holds for the censoring process, i.e., our approach allows censoring to depend in a complex manner on past longitudinal measurements and does not require a model to derive censoring weights. In particular, we feel that there are two general routes to follow for deriving causal effects from observational data; either to make no assumptions for the measurements process but require to derive weighting models for the other competing processes or to make stronger assumptions for the measurement model, and no assumptions for the other competing mechanisms. We have selected here the latter approach.

In this paper, we have focused on the effect of salvage therapy at fixed time $t$ for the individual patient or for a group of patients with similar PSA histories. The framework of joint models could also be used to estimate the causal effect of different policies. For example, one policy might be to start salvage therapy when PSA first goes above 1.0 ng/mL, whereas another policy could delay the start of salvage therapy until PSA first goes above 4.0 ng/mL. A micro-simulation approach could then be used in which the parameter estimates from the joint model allow us to simulate realistic data under both these scenarios, and then the causal effect would be defined as the difference in the incidence rate of metastasis.

As mentioned in Section~\ref{s:UMPD_Analysis}, we have elected only to consider the first time patients received salvage therapy. However, we should stress that this is not a limitation of our proposed modeling framework. Namely, the model could be adjusted to specify the longitudinal subject-specific PSA profiles under multiple salvage therapy interventions. However, we have not done this in our analysis because few patients had received salvage therapy more than once. Hence, there was insufficient information in the data to estimate the changes in the PSA profiles after these interventions.

\section*{Funding}
The authors thank the NIH CISNET Prostate Award CA253910 for financial support.

\section*{Supplementary Material}
Web Figures~1--8, and Web Tables~1--3 are available with this paper as supplementary material.

\begin{spacing}{1}
\twocolumn
\bibliographystyle{smmr}
\bibliography{JM_CI.bib}
\end{spacing}
\end{document}